\documentclass[aps,pre,twocolumn,superscriptaddress]{revtex4}
\usepackage{palatino}
\usepackage{amsmath}
\usepackage{amssymb}
\usepackage{graphicx}
\usepackage{dcolumn}
\usepackage{boxedminipage}
\usepackage{verbatim}
\usepackage[colorlinks=true,citecolor=blue,linkcolor=blue]{hyperref}

\newcommand{\bfv}[1]{{\mbox{\boldmath{$#1$}}}}
\newcommand{\bfm}[1]{{\bf #1}}          

\newcommand{\expect}[1]{\left \langle #1 \right \rangle}                % <.> for denoting expectations over realizations of an experiment or thermal averages

\newcommand{\x}{\bfv{x}}

\newcommand{\bTheta}{\bfm{\Theta}}

\newcommand{\Cov}[1] {\mathrm{cov}\left( #1 \right)}

\newcommand{\T}{\mathrm{T}}                                % T used in matrix transpose

\begin{document}

%% TITLE %%%%%%%%%%%%%%%%%%%%%%%%%%%%%%%%%%%%%%%%%%%%%%%%%%%%%%%%%%%%%%%%%%%%
\title{Statistically optimal analysis of samples from multiple equilibrium states}

\author{Michael R. Shirts}
  \email{michael.shirts@columbia.edu}
 \affiliation{Department of Chemistry, Columbia University, New York, NY 10027}
\author{John D. Chodera}
 \email{jchodera@stanford.edu}
 \affiliation{Department of Chemistry, Stanford University, Stanford, CA 94305}

\date{\today}

%%%%%%%%%%%%%%%%%%%%%%%%%%%%%%%%%%%%%%%%%%%%%%%%%%%%%%%%%%%%%%%%%%%%%%%%%%%%%%%%%%%%%%%%%%%%%%%%%%%%%%
% ABSTRACT/pacs
%%%%%%%%%%%%%%%%%%%%%%%%%%%%%%%%%%%%%%%%%%%%%%%%%%%%%%%%%%%%%%%%%%%%%%%%%%%%%%%%%%%%%%%%%%%%%%%%%%%%%%
\begin{abstract}

% JDC: Revised abstract.
We present a new estimator for computing free energy differences and thermodynamic expectations as well as their uncertainties from samples obtained from multiple equilibrium states via either simulation or experiment.  
The estimator, which we term the multistate Bennett acceptance ratio (MBAR) estimator because it reduces to the Bennett acceptance ratio when only two states are considered, has significant advantages over multiple histogram reweighting methods for combining data from multiple states.  
It does not require the sampled energy range to be discretized to produce histograms, eliminating bias due to energy binning and significantly reducing the time complexity of computing a solution to the estimating equations in many cases.
Additionally, an estimate of the statistical uncertainty is provided for all estimated quantities.  
In the large sample limit, MBAR is unbiased and has the lowest variance of any known estimator for making use of equilibrium data collected from multiple states.  
We illustrate this method by producing a highly precise estimate of the potential of mean force for a DNA hairpin system, combining data from multiple optical tweezer measurements under constant force bias.
\end{abstract}

\maketitle

%%%%%%%%%%%%%%%%%%%%%%%%%%%%%%%%%%%%%%%%%%%%%%%%%%%%%%%%%%%%%%%%%%%%%%%%%%%%%%%%%%%%%%%%%%%%%%%%%%%%%%
% INTRODUCTION
%%%%%%%%%%%%%%%%%%%%%%%%%%%%%%%%%%%%%%%%%%%%%%%%%%%%%%%%%%%%%%%%%%%%%%%%%%%%%%%%%%%%%%%%%%%%%%%%%%%%%%

\section{Introduction}

A recurring challenge in statistical physics, computational chemistry, and single molecule experiments is the collection of a sufficient amount of data to estimate physical quantities of interest to adequate precision.
In computer simulations of physical or chemical models, such quantities include potentials of mean force, phase coexistence curves, fluctuation or temperature-dependent properties, and free energy differences.
In single-molecule experiments, these quantities might include potentials of mean force along a pulling coordinate or the distance between fluorescence probes during resonant energy transfer. 
For all of these problems, collection of sufficient statistics for a reliable estimate often requires multiple simulations at different thermodynamic states~\cite{MLEmeasure_foot1} or measurements performed under different applied biasing potentials.
% JDC: Changed "applied fields" -> "applied biasing potentials" above.
In computer simulations, multi-state techniques such as umbrella sampling~\cite{torrie-valleau:j-comput-phys:1977:umbrella-sampling}, simulated~\cite{marinari-parisi:europhys-lett:1992:simulated-tempering} and parallel tempering~\cite{hukushimi-nemoto:j-phys-soc-jpn:1996:parallel-tempering} and the use of alchemical intermediates in free energy calculations can greatly aid convergence; 
% JDC: Added phrase below.
in experiments, data collected under constant applied force can help provide adequate sampling of conformations of interest~\cite{woodside:prl:2005:optical-force-clamp}.

Even with these methods, it may require a large quantity of data to produce estimates with the desired precision. 
Computing the most precise estimate possible from the available data can therefore be critical in allowing these quantities to be estimated with reasonable computational or experimental effort. 
While the choice of thermodynamic states to sample can also greatly affect efficiency, we focus here on only the problem of statistically efficient estimation given samples from predetermined states.

% JDC: This should be a separate paragraph, since the topic changes!
Early methods for computing free energy differences~\cite{zwanzig:1954a,widom:1963} or equilibrium expectations~\cite{torrie-valleau:j-comput-phys:1977:umbrella-sampling} relied upon one-sided exponential averaging (EXP), which is formally exact but does not make the most efficient use of data when samples from more than one state are available~\cite{shirts:2005a}.
Subsequently, the Bennett acceptance ratio method (BAR)~\cite{bennett:1976,shirts:2003a} greatly improved upon EXP for the computation of free energy differences, producing statistically optimal estimates of free energy differences when two states are sampled~\cite{shirts:2003a} and yielding estimates that can be more than an order of magnitude more precise~\cite{shirts:2005a}.
More recently, multiple histogram reweighting methods~\cite{ferrenberg:1989a,kumar:1992a} were proposed as a way to incorporate data from multiple states to produce superior estimates of free energy differences and equilibrium expectations for arbitrary thermodynamic states, including states not sampled.

While multiple histogram techniques --- most notably, the weighted histogram analysis method (WHAM)~\cite{kumar:1992a} --- can produce statistically optimal estimates of the discretized densities of states~\cite{ferrenberg:1989a} or histogram occupation probabilities~\cite{bartels:1997a}, they have several limitations for the treatment of continuous systems.
First, the reliance on energy histograms of width sufficient to contain many samples --- often larger than many times the thermal energy --- introduces a bias that can be substantial and often difficult to assess~\cite{kobrak:j-comput-chem:2003:histogram-error}.  
% JDC: Added "--- often larger than many kT ---" to above.
Second, unlike BAR, there are no direct expressions to estimate the statistical uncertainty in free energy differences or expectations obtained from WHAM.
Third, application of WHAM to a samples collected with a biasing potential that is not trivially scaled by a linear field parameter
%(such as inverse temperature) 
% JDC: Commented out "(such as inverse temperature)" because it is confusing -- something like an applied electric field along the z-direction might be a simpler example.
requires a number of bins that grows exponentially in the number of states, making it computationally intractable for even modest numbers of states.
While more recent maximum likelihood~\cite{bartels:1997a} and Bayesian formulations~\cite{gallicchio:2005a} mitigate the memory requirements, they do not remove the histogram bias effects, and introduce a costly Markov chain Monte Carlo sampling procedure to estimate uncertainties~\cite{gallicchio:2005a,park:pre:2006:abwham}.

Here, we use recent results from the field of statistical inference~\cite{vardi:1985,gill:1988,kong:2003,tan:2004} to construct a statistically optimal estimator for computing free energy differences and equilibrium expectations at arbitrary thermodynamic states, using equilibrium samples from multiple thermodynamic states.
The resulting estimator, termed the multistate Bennett acceptance ratio (MBAR) estimator as it reduces to Bennett's method when only two states are considered~\cite{bennett:1976}, is equivalent to WHAM in the limit that histogram bin widths are shrunk to zero but is derived without the need to invoke histograms.  
Unlike WHAM, this estimator provides a direct assessment of uncertainties, critical in making comparison between experiment and theory, and the computational expense of computing the estimator remains modest across a wider variety of applications.
Furthermore, it can easily be applied to data sampled from non-Boltzmann sampling schemes, or to the analysis of single-molecule experiments in cases where an external bias potential is applied.

% JDC: Added 
This paper is organized as follows:
Section \ref{section:extended-bridge-sampling} recapitulates the literature on extended bridge sampling estimators, used here as the basis for the MBAR estimator.
Expressions for computing estimates of free energy differences (Section \ref{section:free-energies}) and equilibrium expectations (Section \ref{section:equilibrium-expectations}) are then provided.
Finally, we illustrate the method in Section \ref{section:application} by applying it to the estimation of the potential of mean force for a DNA hairpin system by combining data from multiple equilibrium optical force clamp experiments under different external biasing potentials.

%%%%%%%%%%%%%%%%%%%%%%%%%%%%%%%%%%%%%%%%%%%%%%%%%%%%%%%%%%%%%%%%%%%%%%%%%%%%%%%%%%%%%%%%%%%%%%%%%%%%%%
% EXTENDED BRIDGE SAMPLING ESTIMATION
%%%%%%%%%%%%%%%%%%%%%%%%%%%%%%%%%%%%%%%%%%%%%%%%%%%%%%%%%%%%%%%%%%%%%%%%%%%%%%%%%%%%%%%%%%%%%%%%%%%%%%
\section{Extended bridge sampling estimators}
\label{section:extended-bridge-sampling}

% JDC: Fixed problems here.
Suppose we obtain $N_i$ uncorrelated equilibrium samples from each of $K$ thermodynamic states within the same ensemble, such as NVT, NPT, or $\bfm{\mu}$VT (see Appendix~\ref{appendix:correlated-data} for more information on subsampling correlated timeseries data to produce uncorrelated samples).  
Each state is characterized by a specified combination of inverse temperature, potential energy function, pressure, and/or chemical potential(s), depending upon the ensemble.  
We define the \emph{reduced potential function} $u_i(\x)$ for state $i$ to be
\begin{eqnarray}
u_i(\x) &=& \beta_i [ U_i(\x) + p_i V(\x) + \bfv{\mu}_i^\T \bfv{n}(\x) ] \label{equation:reduced-potential}
\end{eqnarray}
where $\x \in \bfm{\Gamma}$ denotes the configuration of the system within a configuration space $\bfm{\Gamma}$, with volume $V(\x)$ (in the case of a constant pressure ensemble) and $\bfv{n}(\x)$ the number of molecules of each of $M$ components of the system (in the case of a (semi)grand ensemble).  
For each state $i$, $\beta_i$ denotes the inverse temperature, $U_i(\x)$ the potential energy function (which may include biasing weights), $p_i$ the external pressure, and $\bfv{\mu}_i$ the vector of chemical potentials of the $M$ system components.  

Configurations $\{\x_{in}\}_{n=1}^{N_i}$ from state $i$ are sampled from the probability distribution 
\begin{eqnarray}
p_i(\x) = c_i^{-1} q_i(\x) \: &;& \: c_i = \int_{\bfm{\Gamma}} d\x \, q_i(\x)
\end{eqnarray}
where $q_i(\x)$ is here nonnegative and represents an unnormalized density function, and $c_i$ is the (generally unknown) normalization constant (known in statistical mechanics as the \emph{partition function}).
In samples obtained from standard Metropolis Monte Carlo or molecular dynamics simulations or from experiment, this unnormalized density is simply the Boltzmann weight $q_i(\x) = \exp[-u_i(\x)]$ but may in general differ in simulations employing non-Boltzmann weights, such as multicanonical simulations~\cite{mezei:j-comp-phys:1987:muca} and those using Tsallis statistics~\cite{tsallis:j-stat-phys:1988:tsallis-statistics}.

We wish to produce an estimator for the difference in dimensionless free energies
\begin{eqnarray}
\Delta f_{ij} &\equiv& f_j - f_i = - \ln \frac{c_j}{c_i} = - \ln \frac{\int_\bfm{\Gamma} d\x \, q_j(\x)}{\int_\bfm{\Gamma} d\x \, q_i(\x)} \label{equation:dimensionless-free-energy}
\end{eqnarray}
where the $f_i$ are related to the dimensional free energies $F_i$ by $f_i = \beta_i F_i$, and also the equilibrium expectations
\begin{eqnarray}
\expect{A}_i &\equiv& \int_{\bfm{\Gamma}} d\x \, p_i(\x) \, A(\x) = \frac{\int_\bfm{\Gamma} d\x \, A(\x) \, q_i(\x)}{\int_\bfm{\Gamma} d\x \, q_i(\x)} .\label{equation:equilibrium-expectation}
\end{eqnarray}
These expectations can be computed as ratios of the normalization constants if we define new functions $q(\x) = A(\x) q_i(\x)$, where the $q(\x)$ no longer need be nonnegative for states from which no samples are collected~\cite{doss:2003}.

To construct an estimator for these ratios of normalization constants, we first note the identity
\begin{eqnarray}
%c_i \expect{\alpha_{ij} q_j}_i = \int_{\bfm{\Gamma}} d\x \, q_i(\x) \, \alpha_{ij}(\x) \, q_j(\x)  =  c_j \expect{\alpha_{ij} q_i}_j
c_i \expect{\alpha_{ij} q_j}_i &=& \left[\int_{\bfm{\Gamma}} d\x \, q_i(\x)\right] \cdot \frac{\int_{\bfm{\Gamma}} d\x \, q_i(\x) \, \alpha_{ij}(\x) \, q_j(\x)}{\int_{\bfm{\Gamma}} d\x \, q_i(\x)} \nonumber \\
&=& \int_{\bfm{\Gamma}} d\x \, q_i(\x) \, \alpha_{ij}(\x) \, q_j(\x)  \nonumber \\
&=& \left[\int_{\bfm{\Gamma}} d\x \, q_j(\x)\right] \cdot \frac{\int_{\bfm{\Gamma}} d\x \, q_j(\x) \, \alpha_{ij}(\x) \, q_i(\x)}{\int_{\bfm{\Gamma}} d\x \, q_j(\x)} \nonumber \\
&=& c_j \expect{\alpha_{ij} q_i}_j
% JDC: Made steps explicit, since this confused Chris Jarzynski at my poster.
\label{equation:Bennett_starting}
\end{eqnarray}
which holds for arbitrary choice of functions $\alpha_{ij}(\x)$, provided the $c_i$ are nonzero.

Using this relation, summing over the index $j$, and substituting the empirical estimator $N_i^{-1} \sum_{n=1}^{N_i} g(\x_{in})$ for the expectations $\expect{g}_i$, we obtain a set of $K$ estimating equations
\begin{eqnarray}
\sum_{j=1}^K \frac{\hat{c}_i}{N_i} \sum_{n=1}^{N_i} \alpha_{ij} q_j(\x_{in}) &=& \sum_{j=1}^K \frac{\hat{c}_j}{N_j} \sum_{n=1}^{N_j} \alpha_{ij} q_i(\x_{jn}) \label{equation:extended-bridge-sampling-estimator}
\end{eqnarray}
for $i = 1,2,\ldots,K$, where solution of the set of equations for the $\hat{c}_i$ yields an estimate of the $c_i$ from the sampled data determined up to a scalar multiplier.

Eq.~\ref{equation:extended-bridge-sampling-estimator} defines a family of asymptotically unbiased estimators parameterized by the choice of functions $\alpha_{ij}(\x)$, known in the statistics literature as \emph{extended bridge sampling} estimators~\cite{tan:2004}.
By making the choice
\begin{eqnarray}
\alpha_{ij}(\x) &=& N_j \, \hat{c}_j^{-1} / \sum\limits_{k=1}^K N_k \, \hat{c}_k^{-1} \, q_k(\x) \:\:, \label{equation:optimal-alpha}
\end{eqnarray}
we obtain an estimator than has been proven to be optimal, in the sense that it has the lowest variance for a large class of choices of $\alpha_{ij}(\x)$~\cite{tan:2004}.
This estimator is also asymptotically unbiased and guaranteed to have a unique solution (up to a multiplicative constant)~\cite{tan:2004}, and can also been derived from maximum-likelihood measure theoretic methods~\cite{kong:2003} or cast as a reverse logistic regression problem~\cite{geyer:1994,kong:2003}.

While a closed-form expression for the $\hat{c}_i$ cannot be obtained from Eqs.~\ref{equation:extended-bridge-sampling-estimator} and \ref{equation:optimal-alpha}, the $\hat{c}_i$ can nevertheless be easily computed by any suitable method for solving systems of coupled nonlinear equations.
A simple self-consistent iteration method and an efficient Newton-Raphson solver are described in Appendix~\ref{appendix:solution-of-estimating-equations}.

In the large sample limit, the error in the ratios $\hat{c}_i/\hat{c}_j$ will be normally distributed~\cite{tan:2004}, and the asymptotic covariance matrix $\bfm{\Theta}_{ij} = \Cov{\theta_i, \theta_j}$, where $\theta_i \equiv \ln c_i$, can be estimated by~\cite{kong:2003}
\begin{eqnarray}
\hat{\bfm{\Theta}} &=& \bfm{W}^\T (\bfm{I}_N - \bfm{W} \bfm{N} \bfm{W}^\T)^{+} \bfm{W} \label{equation:asymptotic-covariance}
\end{eqnarray}
where $\bfm{I}_N$ is the $N \times N$ identity matrix  (with $N = \sum_{i=1}^K N_i$ the total number of samples), and $\bfm{N} = \mathrm{diag}(N_1, N_2,\ldots,N_K)$.
The superscript $^+$ denotes a suitable generalized inverse, such as the standard Moore-Penrose pseudoinverse, since the quantity in parentheses will be rank-deficient.
$\bfm{W}$ denotes the $N \times K$ matrix of weights
\begin{eqnarray}
W_{ni} = \hat{c}_i^{-1} \frac{q_i(\x_n)}{\sum\limits_{k=1}^K N_k \, \hat{c}_k^{-1} \,\ q_k(\x_n)} \label{equation:p_weights} .
\end{eqnarray}
The samples are now indexed by a single index $n = 1,\ldots,N$, as the association of which samples $\x_n$ came from which distribution $p_i(\x)$ is no longer relevant. 
% JDC: Added back normalization conditions, useful as a check.
We note that this definition ensures $\sum_{n=1}^N  W_{ni} = 1$ for all $i = 1,\ldots,K$ and $\sum_{i=1}^K N_k \, W_{ni} = 1$ for all $n = 1,\ldots,N$.
The computational cost of evaluating the pseudoinverse of an $N \times N$ matrix in computing $\hat{\bfm{\Theta}}$ can be reduced to that of computing the eigenvalue decomposition of a $K \times K$ matrix, and in many cases can be reduced directly to a simpler $K \times K$ matrix (see Appendix~\ref{appendix:efficient-computation-of-asymptotic-covariance}).

The covariance of estimates of arbitrary functions $\phi(\theta_1,\ldots,\theta_K)$ and $\psi(\theta_1,\ldots,\theta_K)$ of the log normalization constants $\theta_i$ can be estimated from $\hat{\bfm{\Theta}}$ by the expansion
\begin{eqnarray}
\Cov{\hat{\phi},\hat{\psi}} &=& \sum_{i,j = 1}^K \frac{\partial \phi}{\partial \theta_i} \,\hat{\Theta}_{ij} \frac{\partial \psi}{\partial \theta_j} \,.\label{equation:asymptotic-covariance-transformation}
\end{eqnarray}

%%%%%%%%%%%%%%%%%%%%%%%%%%%%%%%%%%%%%%%%%%%%%%%%%%%%%%%%%%%%%%%%%%%%%%%%%%%%%%%%%%%%%%%%%%%%%%%%%%%%%%
% ESTIMATOR FOR FREE ENERGY DIFFERENCES
%%%%%%%%%%%%%%%%%%%%%%%%%%%%%%%%%%%%%%%%%%%%%%%%%%%%%%%%%%%%%%%%%%%%%%%%%%%%%%%%%%%%%%%%%%%%%%%%%%%%%%

\section{Free energies}
\label{section:free-energies}

When configurations are sampled with Boltzmann statistics, where $q_i(\x) \equiv \exp[-u_i(\x)]$, Eqs.~\ref{equation:extended-bridge-sampling-estimator} and \ref{equation:optimal-alpha} produce the following estimating equations for the dimensionless free energies
\begin{eqnarray}
\hat f_i &=& - \ln \sum_{j=1}^K \sum_{n=1}^{N_j} \frac{\exp[-u_i(\x_{jn})]}{\sum\limits_{k=1}^K N_k \, \exp[\hat{f}_k - u_k(\x_{jn})]} \label{equation:estimator-of-free-energies}
\end{eqnarray}
which must be solved self-consistency for the $\hat{f}_i$.
Again, because the normalization constants are only determined up to a multiplicative constant, the estimated free energies $\hat{f}_i$ are determined uniquely only up to an additive constant, so only differences $\Delta \hat{f}_{ij} = \hat{f}_j - \hat{f}_i$ will be meaningful.

The uncertainty in the estimated free energy difference can be computed from Eqs.~\ref{equation:asymptotic-covariance} and \ref{equation:asymptotic-covariance-transformation} as
\begin{eqnarray}
\delta^2 \Delta \hat{f}_{ij} &\equiv& \Cov{-\ln \hat{c}_j/\hat{c}_i, -\ln \hat{c}_j/\hat{c}_i} = \hat{\Theta}_{ii} - 2 \hat{\Theta}_{ij} + \hat{\Theta}_{jj} \nonumber
\end{eqnarray}
Free energy differences and uncertainties between states not sampled are easily estimated by augmenting the set of states with additional reduced potentials $u_i(\x)$ with the number of samples $N_i = 0$.  For these unsampled states, no additional self-consistent estimation is required, so many such states can be estimated very efficiently.

%%%%%%%%%%%%%%%%%%%%%%%%%%%%%%%%%%%%%%%%%%%%%%%%%%%%%%%%%%%%%%%%%%%%%%%%%%%%%%%%%%%%%%%%%%%%%%%%%%%%%%
% ESTIMATOR FOR EXPECTATIONS
%%%%%%%%%%%%%%%%%%%%%%%%%%%%%%%%%%%%%%%%%%%%%%%%%%%%%%%%%%%%%%%%%%%%%%%%%%%%%%%%%%%%%%%%%%%%%%%%%%%%%%

\section{Equilibrium expectations}
\label{section:equilibrium-expectations}

The equilibrium expectation of some mechanical observable $A(\x)$ that depends only on configuration $\x$ (and not momentum) is given by Eq.~\ref{equation:equilibrium-expectation}, and can be computed as a ratio of normalization constants $c_A/c_a$ by defining two additional ``states'' characterized by the functions~\cite{doss:2003}
\begin{eqnarray}
q_A(\x) = A(\x) \, q(\x) \:&;&\: q_a(\x) = q(\x) \nonumber 
\end{eqnarray}
where again $q(\x) \equiv \exp[-u(\x)]$ if the expectation with respect to the Boltzmann weight is desired.
Even though $q_A(\x)$ may no longer be strictly nonnegative, we can still make use of the extended bridge sampling estimator (Eq.~\ref{equation:extended-bridge-sampling-estimator}) to estimate the expectation $\expect{A}$ since $N_A = N_a = 0$.

Similarly, we augment the matrix $\bfm{W}$ (Eq.~\ref{equation:p_weights}) with columns $W_{nA}$ and $W_{na}$ corresponding to $q_A(\x)$ and $q_a(\x)$, respectively:
\begin{eqnarray}
W_{nA} &=& \hat{c}_A^{-1} \, \frac{A(\x_n) \, \exp[-u(\x_n)]}{\sum\limits_{k=1}^K N_k \, \exp[\hat{f}_k - u_k(\x_n)]} \nonumber \\
W_{na} &=& \hat{c}_a^{-1} \, \frac{\exp[-u(\x_n)]}{\sum\limits_{k=1}^K N_k \, \exp[\hat{f}_k - u_k(\x_n)]}
\end{eqnarray}
where the normalization constants $\hat{c}_A$ and $\hat{c}_a$ are defined in terms of self-consistent estimating equations as
\begin{eqnarray}
\hat{c}_A &=& \sum\limits_{n=1}^N \frac{A(\x_n) \, \exp[-u(\x_n)]}{\sum\limits_{k=1}^K N_k \, \exp[\hat{f}_k - u_k(\x_n)]} \nonumber \\
\hat{c}_a &=& \sum\limits_{n=1}^N \frac{\exp[-u(\x_n)]}{\sum\limits_{k=1}^K N_k \, \exp[\hat{f}_k - u_k(\x_n)]} 
\end{eqnarray}
We can then write the estimator of the expectation as
\begin{eqnarray}
\hat{A} &=& \frac{\hat{c}_A}{\hat{c}_a} = \sum_{n=1}^N W_{na} \, A(\x_n) \label{equation:estimator-for-expectations} \nonumber
\end{eqnarray}
and an estimator for the uncertainty as
\begin{eqnarray}
\delta^2 \hat{A} \equiv \Cov{\hat{c}_A/\hat{c}_a,\hat{c}_A/\hat{c}_a} = \hat{A}^2 (\hat{\Theta}_{AA} + \hat{\Theta}_{aa} -  2 \, \hat{\Theta}_{Aa}) \nonumber
\end{eqnarray}
where the covariance matrix $\hat{\bfm{\Theta}}$ is now computed from the augmented $\bfm{W}$.
Covariances between estimates of $\expect{A}$ at different thermodynamic states, or between two observables $\expect{A}$ and $\expect{B}$, can also be constructed by adding the appropriate columns to the covariance matrix and applying Eq.~\ref{equation:asymptotic-covariance-transformation} to estimate the desired uncertainty.  
%As previously noted, 
% JDC: Can't find where this was previously noted.
If the dimensionless free energies $\hat{f}_i$ have already been determined, computation of $\hat{A}$ for any $A(\x)$ and any $q(\x)$ does not require additional iterative solution of the self-consistent estimating equations.

%%%%%%%%%%%%%%%%%%%%%%%%%%%%%%%%%%%%%%%%%%%%%%%%%%%%%%%%%%%%%%%%%%%%%%%%%%%%%%%%%%%%%%%%%%%%%%%%%%%%%%
% POTENTIAL OF MEAN FORCE OF DNA HAIRPIN
%%%%%%%%%%%%%%%%%%%%%%%%%%%%%%%%%%%%%%%%%%%%%%%%%%%%%%%%%%%%%%%%%%%%%%%%%%%%%%%%%%%%%%%%%%%%%%%%%%%%%%

\section{Application to laboratory experiments\label{section:application}}

The MBAR estimator is not limited in application to data produced from simulation --- it can also be applied to combine data from multiple equilibrium experiments in the presence of externally-applied fields.
To illustrate, we estimate potential of mean force (PMF) of a DNA hairpin attached by dsDNA linkers to glass beads along the distance between the beads.
The collection of equilibrium trajectories under a variety of constant force loads (corresponding to a linear external potential along the extension coordinate) for the DNA hairpin system 20R55/4T collected by an optical double trap experiment~\cite{woodside:prl:2005:optical-force-clamp} we reported earlier~\cite{woodside:pnas:2006:dna-hairpin-equilibrium}.
The complete dataset was obtained from Michael Woodside (National Institute for Nanotechnology, NRC and Department of Physics, University of Alberta), and consists of 16 trajectories at 296.15 K, each 5 s in duration and sampled with a period of 0.1 ms, totaling 50 000 samples each.
Each trajectory was collected under a different constant force load, with force loads ranging from 12.35 pN to 14.41 pN, with an estimated 10\% relative error in the measurement of this force value.

\begin{figure} 
\caption{\label{figure:pmf} {\bf Potential of mean force of DNA hairpin and dsDNA handles system 20R55/4T under 14.19 pN external force.}} 
\resizebox{\columnwidth}{!}{\includegraphics{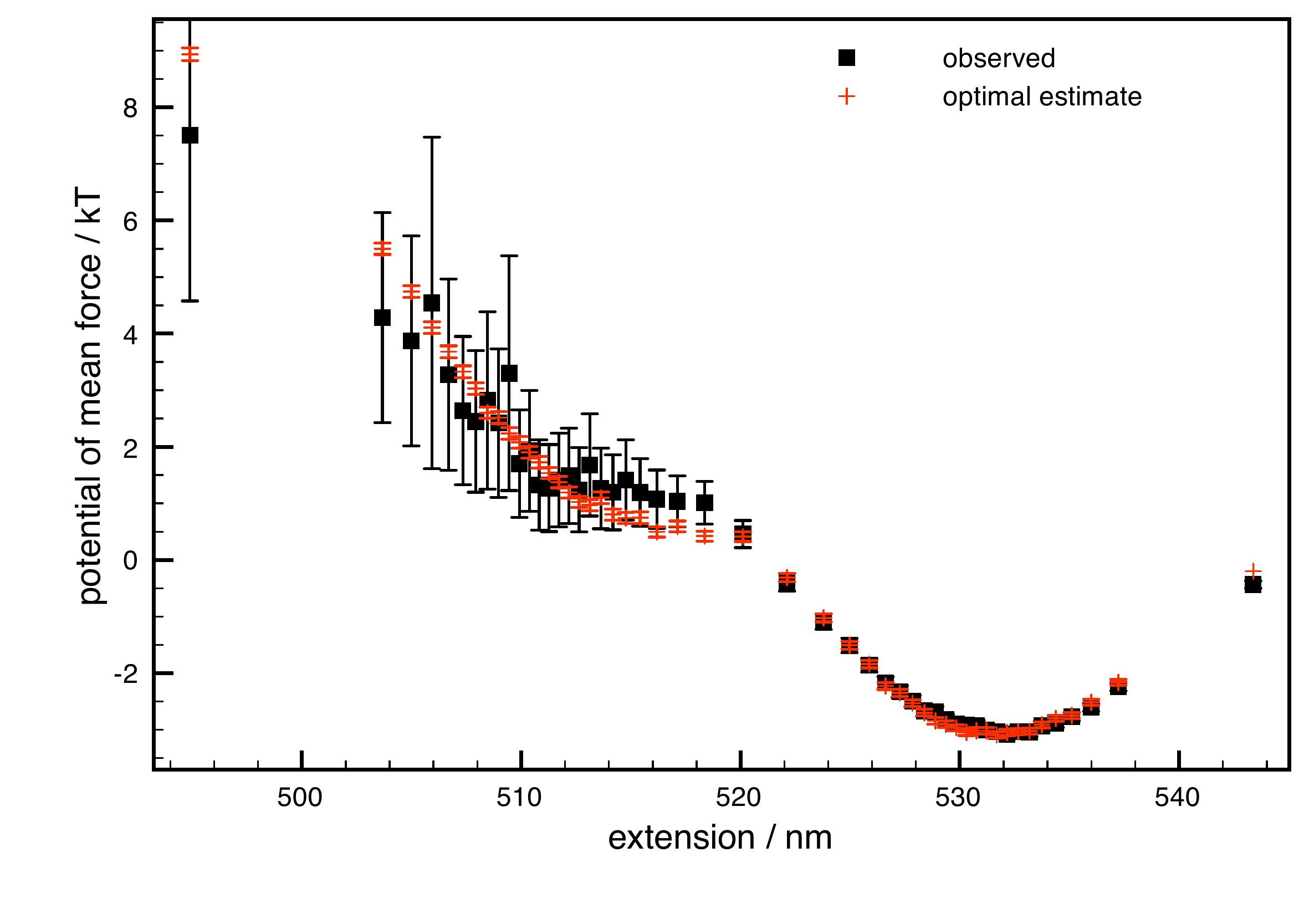}}
\end{figure}

The data were analyzed to produce an optimal estimate of the PMF under a force load of 14.19 pN, a force load where it is difficult to determine the entire PMF to high precision from equilibrium trajectories at this force load alone (Figure~\ref{figure:pmf}).
The sampled extension range was divided up into 50 unequally-sized bins such that the number of samples per bin was equal in order to avoid regions with zero histogram counts, as would occur with equally-spaced bins.
Analysis with the MBAR estimator took 18 s on a standard 2.16 GHz Intel Core 2 Duo MacBook Pro, and the resulting error bars are more than an order of magnitude smaller than those derived from the single trajectory at this force load in the poorly sampled region of the PMF.
Below, we describe how this analysis was performed with both types of analysis.

To estimate the potential of mean force from the 14.19 pN trajectory alone (black filled squares in Figure~\ref{figure:pmf}), the total number of counts $N_i$ per histogram bin was determined, and the reduced potential of mean force (in units of $kT$) $f_i$ computed up to an irrelevant additive constant from
\begin{eqnarray}
f_i &=& - \ln (N_i / w_i)
\end{eqnarray}
where $w_i$ is the relative width of bin $i$, necessary to correct for the nonuniform bin sizes.
The statistical uncertainty in the histogram count was estimated by standard methods (see Eq.~26 of \cite{chodera:jctc:2007})
\begin{eqnarray}
\delta^2 N_i &=& g \, N_i (1 - N_i / N) \nonumber \\
\delta^2 f_i &=& \frac{\delta^2 N_i}{N_i^2}
\end{eqnarray}
where $g$ is the statistical inefficiency of the extension time series, estimated from the extension autocorrelation function (see Section 5.2 of \cite{chodera:jctc:2007}).

To estimate the potential of mean force using the MBAR estimator, the dataset was first subsampled with an interval equal to the statistical inefficiency of each trajectory at constant force to produce a set of uncorrelated samples.
The reduced potential energy for each state $k$ under the experimental conditions corresponds to
\begin{eqnarray}
u_k(\x) &=& \beta [ U_0(\x) + p V(\x) + U^\mathrm{ext}_k(\x)] 
\end{eqnarray}
where $U_0(\x)$ is the (unknown) potential energy function of the system in the absence of an externally-applied biasing potential, $p V(\x)$ is the pressure-volume work, and $U^\mathrm{ext}_k(\x)$ is the (known) externally applied biasing potential, given by
%MRS: do we need to state any assumptions about the pV work?  Can it be neglected here?  We don't really explain anything about it, which might confuse people here. 
\begin{eqnarray}
U^\mathrm{ext}_k(\x) &=& - F_k z(\x) + a_k
\end{eqnarray}
where $z(\x)$ is the extension coordinate, $F_k$ is the constant applied force along the positive $z$ direction, and $a_k$ is a constant offset.

Because only differences $u_i(\x) - u_j(\x)$ appear in the estimating equations (Eq.~\ref{equation:estimator-of-free-energies}), the unknown components of the reduced potential energy cancel out and need not be considered
\begin{eqnarray}
u_i(\x) - u_j(\x) &=& - \beta (F_i - F_j) z(\x) + \beta (a_i - a_j)
\end{eqnarray}
While the constant term $\beta (a_i - a_j)$ involving the unknown zero potential intercepts will appear in the estimated state free energies, these do not influence computed expectations in Eq.~\ref{equation:estimator-for-expectations}, and are hence irrelevant.

The probability of finding the system in a bin $i$ under the conditions of interest is given by the expectation
\begin{eqnarray}
p_i &=& \expect{\chi_i(z)}
\end{eqnarray}
where $\chi_i(z)$ is an indicator function that assumes the value of $1$ if the system is in bin $i$ and zero otherwise.
The potential of mean force (in units of thermal energy $kT$) can then be computed from the $p_i$, up to an irrelevant additive constant, as
\begin{eqnarray}
f_i &=& - \ln (p_i / w_i)
\end{eqnarray}
and the uncertainties propagated by Eq.~\ref{equation:asymptotic-covariance-transformation}.

Because each potential of mean force is only determined up to an arbitrary additive constant, the mean value of each PMF was subtracted before plotting.
This is equivalent to choosing the additive constants so as to obtain an optimal least-squares RMS fit between the two PMFs.

It should be noted that these PMF corresponds to the potential of mean force along the entire system connected to the glass beads, which includes not only the DNA hairpin but the two dsDNA linkers and their attachments to the glass beads.
In other work, deconvolution or related methods have been applied to correct for the stretching of the linkers to estimate PMF for the DNA hairpin alone~\cite{woodside:science:2006:dna-hairpin-optical-trap}.

%%%%%%%%%%%%%%%%%%%%%%%%%%%%%%%%%%%%%%%%%%%%%%%%%%%%%%%%%%%%%%%%%%%%%%%%%%%%%%%%%%%%%%%%%%%%%%%%%%%%%%
% DISCUSSION
%%%%%%%%%%%%%%%%%%%%%%%%%%%%%%%%%%%%%%%%%%%%%%%%%%%%%%%%%%%%%%%%%%%%%%%%%%%%%%%%%%%%%%%%%%%%%%%%%%%%%%

\section{Discussion}

The MBAR estimator presented here provides a rapid and robust way to extract estimates of free energy differences and equilibrium expectations from multiple equilibrium samples of different thermodynamic states in a statistically optimal way.
As the estimator is asymptotically efficient among a wide class of ``bridge sampling'' estimators~\cite{tan:2004}, which includes EXP and BAR as members, the resulting estimates from MBAR will have the lowest (or equal) variance in the large sample limit.

While multiple histogram techniques~\cite{kumar:1992a,ferrenberg:1989a,bartels:1997a,gallicchio:2005a} have been widely used for combining data from multistate simulations, the MBAR estimator supplants these methods in the majority of cases.
Most importantly, it provides a reliable and inexpensive method for estimating the uncertainties in the resulting estimates and their correlations, which are critical for propagating uncertainties to quantities of interest.  
Additionally, the elimination of histograms avoids both the bias arising from discretization of continuous energies as well as the computational overhead of constructing and storing high dimensional histograms.  

In this framework, multiple histogram reweighting methods such as WHAM can be understood as a histogram kernel density estimator approximation to MBAR.
In some applications, histograms can reduce the computational expense required for solving the estimating equations (Eq.~\ref{equation:extended-bridge-sampling-estimator}) at the expense of introducing bias. 
When samples are distributed according to the Boltzmann weight, the estimator for the free energies (Eq.~\ref{equation:estimator-of-free-energies}) is precisely Eq.~21 of~\cite{kumar:1992a} or Eq.~15 of~\cite{souaille:2001a}, in both cases presented as a reduction of the histogram bin width to zero in the standard WHAM equations (Eqs.~19--20 of \cite{kumar:1992a}).
While the validity of this limit is dubious --- the derivations in these references rely upon an estimate of the uncertainty in each histogram count which cannot be correct when the bins are nearly empty --- the derivation of this equation from the extended bridge sampling estimator demonstrates for the first time that these equations are, in fact, asymptotically unbiased estimators of the true free energy differences.  

The MBAR estimator also can be considered a multistate generalization of Bennett's acceptance ratio estimator (BAR)~\cite{bennett:1976}.  
In deriving BAR, Bennett constructed an estimator from Eq.~\ref{equation:Bennett_starting} directly, determining the single $\alpha(\x)$ which minimized the variance of the estimator of the free energy difference between only two states.
In deriving MBAR, summing over all states $j$ and determining the functions $\alpha_{ij}(\x)$ that minimizes the covariance matrix of the estimator for ratios of normalization constants produces an optimal estimator for the multistate case.  
A proof of the equivalence of MBAR and BAR for two states can be found in Appendix~\ref{appendix:equivalence-to-bar}.

BAR and a recent pairwise multistate generalization (which we shall refer to as PBAR)~\cite{maragakis:prl:2006} differ from MBAR in that they can also be applied to \emph{nonequilibrium} work measurements between pairs of states, in addition to equilibrium reduced potential differences (\emph{instantaneous} work measurements).  
However, PBAR constructs a total likelihood function from products of likelihood functions connecting pairs of states, assuming independence of all work measurements.  
For equilibrium samples, this means that a sampled configuration $\x_{n}$ from a state $i$ can only be used to provide information about the instantaneous work required to switch to a \emph{single} other state $j$ for use in the PBAR estimator, whereas in MBAR, each sampled $\x_{n}$ can be used to provide information about \emph{all} states.  
As a result, MBAR should require significant fewer samples from each state to produce an estimate of equivalent precision with equilibrium data.

A Python implementation of the MBAR estimator described here is available under the GNU General Public License (GPL), and is provided online, along with several example applications, at \url{https://simtk.org/home/pymbar}.

%%%%%%%%%%%%%%%%%%%%%%%%%%%%%%%%%%%%%%%%%%%%%%%%%%%%%%%%%%%%%%%%%%%%%%%%%%%%%%%%%%%%%%%%%%%%%%%%%%%%%%
% ACKNOWLEDGMENTS
%%%%%%%%%%%%%%%%%%%%%%%%%%%%%%%%%%%%%%%%%%%%%%%%%%%%%%%%%%%%%%%%%%%%%%%%%%%%%%%%%%%%%%%%%%%%%%%%%%%%%%

\section{Acknowledgments}

We are indebted to Michael Woodside for providing us with detailed datasets from Ref.~\cite{woodside:pnas:2006:dna-hairpin-equilibrium} and helpful comments.
We thank Evangelos A.~Coutsias, Gavin E.~Crooks, Fernando A.~Escobedo, Edward H.~Feng, Andrew Gelman, Andrew I.~Jewett, Libusha Kelly, Jun S.~Liu, David D.~L.~Minh,  David L.~Mobley, Frank M.~No\'{e}, Vijay S.~Pande, Sanghyun Park, M.~Scott Shell, Zhiqiang Tan, Matthew A.~Wyczalkowski, and the anonymous reviewers for enlightening discussions and constructive comments on this manuscript. 
JDC gratefully acknowledges support from Ken A.~Dill through NIH grant GM34993 and Vijay S.~Pande through an NSF grant for Cyberinfrastructure (NSF CHE-0535616), and MRS support from Richard A.~Friesner and an NIH NRSA Fellowship.

%%%%%%%%%%%%%%%%%%%%%%%%%%%%%%%%%%%%%%%%%%%%%%%%%%%%%%%%%%%%%%%%%%%%%%%%%%%%%%%%%%%%%%%%%%%%%%%%%%%%%%
%%%%%%%%%%%%%%%%%%%%%%%%%%%%%%%%%%%%%%%%%%%%%%%%%%%%%%%%%%%%%%%%%%%%%%%%%%%%%%%%%%%%%%%%%%%%%%%%%%%%%%
% APPENDIX
%%%%%%%%%%%%%%%%%%%%%%%%%%%%%%%%%%%%%%%%%%%%%%%%%%%%%%%%%%%%%%%%%%%%%%%%%%%%%%%%%%%%%%%%%%%%%%%%%%%%%%
%%%%%%%%%%%%%%%%%%%%%%%%%%%%%%%%%%%%%%%%%%%%%%%%%%%%%%%%%%%%%%%%%%%%%%%%%%%%%%%%%%%%%%%%%%%%%%%%%%%%%%
\appendix
\begin{widetext}

%%%%%%%%%%%%%%%%%%%%%%%%%%%%%%%%%%%%%%%%%%%%%%%%%%%%%%%%%%%%%%%%%%%%%%%%%%%%%%%%%%%%%%%%%%%%%%%%%%%%%%
% SOLUTION OF THE ESTIMATING EQUATIONS
%%%%%%%%%%%%%%%%%%%%%%%%%%%%%%%%%%%%%%%%%%%%%%%%%%%%%%%%%%%%%%%%%%%%%%%%%%%%%%%%%%%%%%%%%%%%%%%%%%%%%%

\section{Correlated time series data\label{appendix:correlated-data}}

While estimating equations based on Eq.~\ref{equation:extended-bridge-sampling-estimator} can be applied to correlated or uncorrelated datasets, provided that the empirical estimator
\begin{eqnarray}
\expect{A}_i \approx \frac{1}{N_i} \sum_{n=1}^{N_i} A(\x_{in})
\end{eqnarray}
remains asymptotically unbiased, the asymptotic covariance matrix estimator (Eq.~\ref{equation:asymptotic-covariance}) only produces sensible estimates when applied to \emph{uncorrelated} datasets.
Application to correlated datasets may produce severe underestimates of the true statistical uncertainty, and should be avoided.

A set of uncorrelated configurations can be obtained from a correlated time series, such as is generated by a molecular dynamics or Metropolis Monte Carlo simulation, by subsampling the time series with an interval approximately equal to the equilibrium relaxation time for the system.
Because the equilibrium relaxation time is difficult to compute for all but the simplest systems, we find the \emph{maximum} of the statistical inefficiency $g$ computed for several relevant observables (such as the reduced potential $u_k(\x)$ in Boltzmann-weighted sampling, structural observables $A(\x)$ in the computation of potentials of mean force, etc.) provides a practical estimate useful for subsampling.

The statistical inefficiency $g_A$ of the observable $A(\x)$ of a time series $\{\x_t\}_{t=1}^T$ is formally defined as (see Janke~\cite{janke:2002a} for a detailed exposition)
\begin{eqnarray}
g_A &\equiv& 1 + 2 \, \tau_A \nonumber \\
\tau_A &\equiv& \sum_{t = 1}^T \left(1 - t/T\right) \, C_{AA}(t) \nonumber \\
C_{AA}(t) &\equiv& \frac{\expect{A(\x_{t_0}) \, A(\x_{t_0+t})}_i - \expect{A}_i^2}{\expect{A^2}_i - \expect{A}_i^2}
\end{eqnarray}
where $\tau_A$ denotes the integrated autocorrelation time and $C_{AA}(t)$ the normalized fluctuation autocorrelation function of the observable $A$.

Direct application of these equations substituting the empirical estimator for the expectation can be problematic due to statistical noise.
As a result, there exist a number of standard procedures~\cite{swope:1982a,flyvbjerg:1989a,janke:2002a,chodera:jctc:2007} to improve the quality and stability of this estimate for physical systems, making use of properties such as stationarity.
The fast method for estimating the integrated autocorrelation time described in Section 5.2 of Chodera et al.~\cite{chodera:jctc:2007} is implemented in the Python implementation of MBAR available online.

%%%%%%%%%%%%%%%%%%%%%%%%%%%%%%%%%%%%%%%%%%%%%%%%%%%%%%%%%%%%%%%%%%%%%%%%%%%%%%%%%%%%%%%%%%%%%%%%%%%%%%
% RECOMPUTATION OF ENERGIES AT MULTIPLE STATES
%%%%%%%%%%%%%%%%%%%%%%%%%%%%%%%%%%%%%%%%%%%%%%%%%%%%%%%%%%%%%%%%%%%%%%%%%%%%%%%%%%%%%%%%%%%%%%%%%%%%%%

\section{Recomputation of reduced energies at multiple states\label{appendix:recomputation}}

Application of MBAR to simulation data requires $u_k(\x_n)$ to be evaluated for all $K$ reduced potential functions $u_k(\x)$ and all $N$ uncorrelated sampled configuration $\x_{n}$, a total of $KN$ reduced potential evaluations. 
In practice, this is not overly burdensome; the samples $\x_{n}$ are generally produced by schemes that generate chains of highly correlated samples, such as molecular dynamics or Monte Carlo simulations. 
Once the stored configurations are subsampled to eliminate correlations and produce an effectively uncorrelated sample (as described above), the number of remaining samples $N \approx T / g$ is generally smaller than the number of samples $T$ produced during the simulation by one or more orders of magnitude.

In cases where the $u_k(\x)$ differ only by a linear scaling parameter of one or more components (such as temperature or an external field parameter), computation of $u_k(\x_n)$ for all $K$ states is a trivial operation.
For other cases, such as when all samples are collected from thermodynamic states that only differ in the external biasing potential (e.g.~linear or harmonic), we note that the reduced energy differences $u_k(\x) - u_i(\x)$ involve only differences in the external biasing potential, which can often be rapidly computed.
Section~\ref{section:application} contains an illustration of this in application to single-molecule pulling experiments.

\section{Efficient solution of the estimating equations\label{appendix:solution-of-estimating-equations}}

A number of methods can be used to obtain a self-consistent solution to the free energy estimating equations obtained from combining Eqs.~\ref{equation:extended-bridge-sampling-estimator} and \ref{equation:optimal-alpha}
\begin{eqnarray}
\hat{c}_i &=& \sum_{j=1}^K \sum_{n=1}^{N_j} \frac{q_i(\x_{jn})}{\sum\limits_{k=1}^K N_k \, \hat{c}_k^{-1} \, q_k(\x_{jn})} \label{equation:optimal-estimator}.
\end{eqnarray}
or, in terms of the dimensionless free energies $f_i = - \ln c_i$,
\begin{eqnarray}
\hat f_i &=& - \ln \sum_{j=1}^K \sum_{n=1}^{N_j} \frac{q_i(\x_{jn})}{\sum\limits_{k=1}^K N_{k} \, e^{\hat{f}_k} \, q_k(\x_{jn})} \label{equation:appendix-estimating-equations} .
\end{eqnarray}
While any method capable of solving a coupled set of nonlinear equations may be employed here, we describe some practical choices we made in the implementation of this algorithm.
While any method capable of solving a coupled set of nonlinear equations may be employed, we describe two approaches to their solution: a straightforward yet reliable self-consistent iteration method and an efficient yet slightly less reliable Newton-Raphson method.
Both methods are implemented in the Python implementation of the estimator available online.

\subsection{Self-consistent iteration\label{section:self-consistent-iteration}}
As in \cite{kumar:1992a}, the $\hat{f}_i$ could be obtained by self-consistent iteration of Eq.~\ref{equation:estimator-of-free-energies} using the last set of iterates $\{\hat{f}_i^{(n)}\}_{i=1}^K$ to produce a new estimated set of iterates $\{\hat{f}_i^{(n+1)}\}_{i=1}^K$:
\begin{eqnarray}
\hat f_i ^{(n+1)} &=& - \ln \sum_{j=1}^K \sum_{n=1}^{N_j} \frac{q_i(\x_{jn})}{\sum\limits_{k=1}^K N_k \, e^{\hat{f}_k^{(n)}} \, q_k(\x_{jn})} .
\end{eqnarray}
Convergence is assured regardless of the initial choice of $f_i^{(0)}$, so it is sufficient to initialize the iteration by choosing $f_i^{(0)} = 0$.
Alternative initial choices of the initial reduced free energies $f_i^{(0)}$ may speed convergence.
For example, we have found the choice
\begin{eqnarray}
f_k^{(0)} &=& \frac{1}{N_k} \sum_{n=1}^{N_k} \ln q(\x_{kn})
\end{eqnarray}
which, for Boltzmann weighting ($q_k(\x) = \exp[-u_k(\x)]$) corresponds to the average reduced potential energy, usually works well.
Additional inexpensive choices are possible, such as fixing $f_1^{(0)} = 0$ and estimating consecutive differences $(f_{k+1}^{(0)} - f_{k}^{(0)})$, $k = 1,2,\ldots,K-1$ using the Bennett acceptance ratio (BAR) estimator~\cite{bennett:1976,shirts:2003a}.

\subsubsection{Cautions}

For numerical reasons, it is convenient to constrain $f_1 = 0$ during the course of iteration by subtracting $f_1$ from the updated values in order to obtain a unique solution and prevent uncontrolled growth in the magnitude of the estimates.
Iteration is terminated when the quantities of interest change by a fraction of the desired precision with additional iterations, but a convenient rule of thumb is to terminate when $\max_{i = 2,\ldots,K} |f_i^{(n+1)} - f_i^{(n)}| / |f_i^{(n)}| < 10^{-7}$. 
Because the quantities of interest and the relative free energies can converge at different rates, it is advised that the former be monitored when possible.

It is also critical to avoid \emph{overflow} in the computation of exponentials $e^a$.
To compute log sums of the form $\ln \sum_{n=1}^N e^{a_n}$, we can use the equivalent form
\begin{eqnarray}
\ln \sum_{n=1}^N \exp[a_n] = c + \ln \sum_{n=1}^N \exp[a_n - c]
\end{eqnarray}
where $c \equiv \max_{n} a_n$.
To minimize \emph{underflow}, the terms $\exp[a_n - c]$ can be summed in order from smallest to largest.

\subsection{Newton-Raphson}

A more efficient approach to determination of the $\hat{f}_i$ is to employ a Newton-Raphson solver, which has the advantage of quadratic convergence (a near doubling of the number of digits of precision) with each iteration when sufficiently near the solution.
Because each iteration requires inversion of a $(K-1) \times (K-1)$ matrix, this approach is only efficient if $K$ is small, say $K < 100$, but this will be satisfied in a wide number of cases.

First, we write the estimating equations in terms of a set of functions $g_i(\bfv{\theta})$ such that the solution of the estimating equations (Eq.~\ref{equation:appendix-estimating-equations}) corresponds to $\bfm{g}(\hat{\bfv{\theta}}) = \bfm{0}$.
Several such choices of both the function $\bfm{g}(\bfv{\theta})$ and the parameterization (the normalization constants $c_i$ or their logarithms $\theta_i$) are possible, and the efficiencies of approaches based on different choices may differ substantially, but we find it convenient to choose
\begin{eqnarray}
g_i(\bfv{\theta}) &=& N_i - N_i \sum_{n=1}^N W_{ni}(\bfv{\theta}) .
\end{eqnarray}
where $W_{ni}$ is defined in Eq.~\ref{equation:p_weights}.
It can easily be seen that $\bfm{g}(\hat{\bfv{\theta}}) = \bfm{0}$ is equivalent to the estimating equations:
\begin{eqnarray}
&& g_i(\hat{\bfv{\theta}}) = 0 \nonumber \\
\Leftrightarrow \:\: && N_i - N_i \sum\limits_{n=1}^N W_{ni}(\hat{\bfv{\theta}}) = 0 \nonumber \\
\Leftrightarrow \:\: && N_i  - N_i \frac{\hat{c}_i^{-1} \, q_i(\x_n)}{\sum\limits_{k=1}^K N_k \, \hat{c}_k^{-1} \,\ q_k(\x_n)} = 0 \nonumber \\
\Leftrightarrow \:\: && \hat{c}_i = \frac{q_i(\x_n)}{\sum\limits_{k=1}^K N_k \, \hat{c}_k^{-1} \,\ q_k(\x_n)}
\end{eqnarray}

In Newton-Raphson, the function $\bfm{g}(\bfv{\theta})$ is expanded about the current iterate $\bfv{\theta}^{(n)}$ to first order:
\begin{eqnarray}
\bfm{g}(\bfv{\theta}^{(n+1)}) &\approx& \bfm{g}(\bfv{\theta}^{(n)}) + \bfm{H}(\bfv{\theta}^{(n)}) (\bfv{\theta}^{(n+1)} - \bfv{\theta}^{(n)}) . \label{equation:newton-raphson-taylor-expansion}
\end{eqnarray}
where 
\begin{eqnarray}
H_{ij}(\bfv{\theta}) &=& \frac{\partial g_i(\bfv{\theta})}{\partial \theta_j}
= \begin{cases}
- \sum\limits_{n=1}^N N_i \, W_{ni} ( 1 - N_i \, W_{ni} ) & \mathrm{if} \:\: i = j \\
\sum\limits_{n=1}^N N_i \,W_{ni} \, N_j \, W_{nj} & \mathrm{if} \:\: i \ne j
\end{cases}
\end{eqnarray}
We seek the next iterate $\bfv{\theta}^{(n+1)})$ such that $\bfm{g}(\bfv{\theta}^{(n+1)}) = \bfm{0}$, which yields the update equation
\begin{eqnarray}
\bfv{\theta}^{(n+1)} &=& \gamma \, [\bfm{H}(\bfv{\theta}^{(n)})]^+ \bfm{g}(\bfv{\theta}^{(n)}) \label{equation:newton-raphson-iteration}
\end{eqnarray}
where $^+$ denotes the pseudoinverse.
If all the $q_i(\x)$ are unique and $N_i > 0$ for all states, the standard matrix inverse may be substituted for the pseudoinverse.

\subsubsection{Cautions}

We only need to iterate over states for which $N_i > 0$; the relative free energies of states where $N_i = 0$, and expectation values at all states, can be determined after the self-consistent equations are solved to determine the relative free energies of states where $N_i > 0$.
Since we must constrain $f_1 = 0$ to avoid drift during the process of free energy determination, we can simply use a modified form of Eq.~\ref{equation:newton-raphson-iteration} where rows and columns corresponding to the first state are omitted.
\begin{eqnarray}
\bfv{\theta}^{(n+1)}_{(2:K,2:K)} &=& \gamma \, [\bfm{H}(\bfv{\theta}^{(n)})_{(2:K,2:K)}]^+ \bfm{g}(\bfv{\theta}^{(n)})_{(2:K,2:K)} . \label{equation:newton-raphson-iteration-truncate}
\end{eqnarray}

$\gamma \in (0,1]$ is a scalar multiplier that controls the rate of convergence.
Since the initial iterate $\bfv{\theta}^{(1)}$ may be far from the realm of quadratic convergence (i.e.~outside the range at which the Taylor expansion in Eq.~\ref{equation:newton-raphson-taylor-expansion} holds), it is often safer to choose an initial $\gamma \ll 1$.
We have found $\gamma = 0.1$ works well for the first step, with $\gamma = 1$ used thereafter.

Even then, there are times when  with reduced $\gamma$ does not prevent numerical instability.
The instability may be due to the initial guess iterate $\bfv{\theta}^{(0)}$ being too far from the region of quadratic convergence, such that the first-order Taylor expansion above is a poor approximation to $\bfm{g}(\bfv{\theta})$ in Eq.~\ref{equation:newton-raphson-taylor-expansion}.
In this case, a better procedure for choosing the initial iterate may aid convergence.
Starting with one or more iterations of the self-consistent method (Section \ref{section:self-consistent-iteration}), or using an initial estimate from application of BAR~\cite{shirts:2003a} to sequential states may be sufficient. 
Less commonly, failure to converge may be result from numerical precision limiting the accurate calculation of the pseudoinverse $[\bfm{H}(\bfv{\theta}^{(n)})_{(2:K,2:K)}]^+$.
In all cases, we find that self-consistent iteration still works reliably to recover the estimator, and can be used as a fallback procedure.

%%%%%%%%%%%%%%%%%%%%%%%%%%%%%%%%%%%%%%%%%%%%%%%%%%%%%%%%%%%%%%%%%%%%%%%%%%%%%%%%%%%%%%%%%%%%%%%%%%%%%%
% COMPUTING THE COVARIANCE MATRIX
%%%%%%%%%%%%%%%%%%%%%%%%%%%%%%%%%%%%%%%%%%%%%%%%%%%%%%%%%%%%%%%%%%%%%%%%%%%%%%%%%%%%%%%%%%%%%%%%%%%%%%

\section{Efficient computation of the asymptotic covariance matrix\label{appendix:efficient-computation-of-asymptotic-covariance}}

\subsection{Singular value decomposition}

The $N \times K$ matrix $\bfm{W}$ (Eq.~\ref{equation:p_weights} in the main paper) can be written in terms of its singular value decomposition
\begin{eqnarray}
\bfm{W} &=& \bfm{U} \bfm{\Sigma} \bfm{V}^\T
\end{eqnarray}
where $\bfm{U}$ is an $N \times N$ unitary matrix of left singular vectors (such that $\bfm{U} \bfm{U}^\T = \bfm{I}_N$), $\bfm{\Sigma}$ is an $N \times K$ matrix containing $L < K$ singular values along the diagonal, and $\bfm{V}$ is a $K \times K$ unitary matrix of right singular vectors. 

The estimator for the asymptotic covariance matrix $\hat{\bfm{\Theta}}$ (Eq.~\ref{equation:asymptotic-covariance}) can then be expanded to
%(Eq. 6 in the main paper) 
% JDC: Let's worry about separating the main body and supplementary info only at the final, final stage before submission.
\begin{eqnarray}
\hat{\bfm{\Theta}} &=& \bfm{W}^\T (\bfm{I}_N - \bfm{W} \bfm{N} \bfm{W}^\T )^+ \bfm{W} \nonumber \\
&=& (\bfm{U} \bfm{\Sigma} \bfm{V}^\T)^\T [\bfm{I}_N - (\bfm{U} \bfm{\Sigma} \bfm{V}^\T) \bfm{N} (\bfm{U} \bfm{\Sigma} \bfm{V}^\T)^\T ]^+ (\bfm{U} \bfm{\Sigma} \bfm{V}^\T) \nonumber \\
&=& \bfm{V} \bfm{\Sigma}^\T \bfm{U}^\T [\bfm{I}_N - \bfm{U} \bfm{\Sigma} \bfm{V}^\T \bfm{N} \bfm{V} \bfm{\Sigma}^\T \bfm{U}^\T ]^+ \bfm{U} \bfm{\Sigma} \bfm{V}^\T \nonumber \\
&=& \bfm{V} \bfm{\Sigma}^\T \bfm{U}^\T [\bfm{U} \bfm{U}^\T - \bfm{U} \bfm{\Sigma} \bfm{V}^\T \bfm{N} \bfm{V} \bfm{\Sigma}^\T \bfm{U}^\T ]^+ \bfm{U} \bfm{\Sigma} \bfm{V}^\T \nonumber \\
&=& \bfm{V} \bfm{\Sigma}^\T \bfm{U}^\T [\bfm{U} ( \bfm{I}_N - \bfm{\Sigma} \bfm{V}^\T \bfm{N} \bfm{V} \bfm{\Sigma}^\T )\bfm{U}^\T ]^+ \bfm{U} \bfm{\Sigma} \bfm{V}^\T \nonumber \\
&=& \bfm{V} \bfm{\Sigma}^\T \bfm{U}^\T \bfm{U} [\bfm{I}_N - \bfm{\Sigma} \bfm{V}^\T \bfm{N} \bfm{V} \bfm{\Sigma}^\T ]^+ \bfm{U}^\T \bfm{U} \bfm{\Sigma} \bfm{V}^\T \nonumber \\
&=& \bfm{V} \bfm{\Sigma}^\T [\bfm{I}_N - \bfm{\Sigma} \bfm{V}^\T \bfm{N} \bfm{V} \bfm{\Sigma}^\T ]^+ \bfm{\Sigma} \bfm{V}^\T \nonumber \\
\end{eqnarray}
We partition the matrix of singular values $\bfm{\Sigma}$ into a $K \times K$ diagonal region $\Sigma_K$ (of which only the first $L \le K$ diagonal entries will be nonzero) and an $(N-K) \times K$ zero matrix $\bfm{0}$:
\begin{eqnarray}
\bfm{\Sigma} &=& 
\left[ \begin{array}{c}
\bfm{\Sigma}_K \\
\bfm{0}
\end{array}
\right]
\end{eqnarray}
We can then rewrite the above expression as
\begin{eqnarray}
\hat{\bfm{\Theta}} &=& \bfm{V} 
\left[ \begin{array}{cc}
\bfm{\Sigma}_K & \bfm{0}
\end{array}
\right]
\left\{
\left[ \begin{array}{cc}
\bfm{I}_K & \bfm{0}\\
\bfm{0} & \bfm{I}_{(N-K)}
\end{array}
\right]
- 
\left[ \begin{array}{c}
\bfm{\Sigma}_K \\
\bfm{0}
\end{array}
\right]
\bfm{V}^\T \bfm{N} \bfm{V} 
\left[ \begin{array}{cc}
\bfm{\Sigma}_K & \bfm{0}
\end{array}
\right]
\right\}^+ 
\left[ \begin{array}{c}
\bfm{\Sigma}_K \\
\bfm{0}
\end{array}
\right]
\bfm{V}^\T \nonumber \\
&=& \bfm{V} \bfm{\Sigma}_K [\bfm{I}_K - \bfm{\Sigma}_K \bfm{V}^\T \bfm{N} \bfm{V} \bfm{\Sigma}_K ]^+ \bfm{\Sigma}_K \bfm{V}^\T \label{equation:asymptotic-covariance-in-terms-of-svd}
\end{eqnarray}
We note that pseudoinversion of the quantity in brackets now only requires $\mathcal{O}(K^3)$ work, though this can be further reduced to $\mathcal{O}(L^3)$ work if the reduced SVD is used.

The singular values $\bfm{\Sigma}_K$ and matrix of right singular vectors $\bfm{V}$ can easily be computed from the eigenvalue decomposition of $\bfm{W^\T W}$:
\begin{eqnarray}
\bfm{W}^\T \bfm{W} &=& (\bfm{U} \bfm{\Sigma} \bfm{V}^\T)^\T (\bfm{U} \bfm{\Sigma} \bfm{V}^\T) \nonumber \\
&=& \bfm{V} \bfm{\Sigma}^T \bfm{U}^\T \bfm{U} \bfm{\Sigma} \bfm{V}^\T \nonumber \\
&=& \bfm{V} (\bfm{\Sigma}^T \bfm{\Sigma}) \bfm{V}^\T
\end{eqnarray}

\subsection{When $\bfm{W}$ has full column rank}

In the case that $\bfm{W}$ has full column rank (because all $q_k(\x)$, $k = 1,\ldots,K$ are unique) we can make further progress.
Using Eq.~\ref{equation:asymptotic-covariance-in-terms-of-svd}, we can write
\begin{eqnarray}
\hat{\bfm{\Theta}} &=& [ \bfm{V} \bfm{\Sigma}_K^{-1} ( \bfm{I}_K - \bfm{\Sigma}_K \bfm{V}^\T \bfm{N} \bfm{V} \bfm{\Sigma}_K + \bfm{1} \bfm{1}^\T ) \bfm{\Sigma}_K^{-1} \bfm{V}^\T ]^{-1} \nonumber \\
 &=& [ \bfm{V} (\bfm{\Sigma}_K^{-2}) \bfm{V}^\T - \bfm{N} ]^+ \nonumber \\
 &=& [ (\bfm{W}^\T \bfm{W})^{-1} - \bfm{N} ]^+
\end{eqnarray}
We note that $\bfm{W}^\T \bfm{1}_N = \bfm{1}_K$, and $\bfm{W} \bfm{N} \bfm{1}_K = \bfm{1}_N$, and so $\bfm{W}^\T \bfm{W} \bfm{N} \bfm{1}_N = \bfm{1}_K$, and observe that $[(\bfm{W}^\T \bfm{W})^{-1} - \bfm{N}]$ has rank $K-1$ with kernel $\bfm{1}_K$
\begin{eqnarray}
[(\bfm{W}^\T \bfm{W})^{-1} - \bfm{N}] \bfm{1}_K &=& (\bfm{W}^\T \bfm{W})^{-1} \bfm{1}_K - \bfm{N} \bfm{1}_K \nonumber \\
&=& (\bfm{W}^\T \bfm{W})^{-1} \bfm{W}^\T \bfm{W} \bfm{N} \bfm{1}_K - \bfm{N} \bfm{1}_K \nonumber \\
&=& \bfm{N} \bfm{1}_K - \bfm{N} \bfm{1}_K \nonumber \\
&=& \bfm{0} \nonumber
\end{eqnarray}
We can supplement the quantity in brackets with $b \bfm{1}_K \bfm{1}_K^\T$, where $b$ is some nonzero scalar, without changing the covariance values computed from it, and make it invertible:
\begin{eqnarray}
\hat{\bfm{\Theta}} &=& [ (\bfm{W}^\T \bfm{W})^{-1} - \bfm{N} + b \bfm{1}_K \bfm{1}_K ]^{-1} \label{equation:covariance-matrix-full-rank}
\end{eqnarray}
We choose $b = N^{-1}$ to ensure the inversion is well-conditioned (as in \cite{kong:2003}), producing
\begin{eqnarray}
\hat{\bfm{\Theta}} &=& [(\bfm{W}^\T \bfm{W})^{-1} -\bfm{N} + \bfm{1}_K \bfm{1}_K^{\T}/N]^{-1} \label{equation:simplified-asymptotic-covariance}.
\end{eqnarray}

%%%%%%%%%%%%%%%%%%%%%%%%%%%%%%%%%%%%%%%%%%%%%%%%%%%%%%%%%%%%%%%%%%%%%%%%%%%%%%%%%%%%%%%%%%%%%%%%%%%%%%
% RELATIONSHIPS BETWEEN BAR AND MBAR
%%%%%%%%%%%%%%%%%%%%%%%%%%%%%%%%%%%%%%%%%%%%%%%%%%%%%%%%%%%%%%%%%%%%%%%%%%%%%%%%%%%%%%%%%%%%%%%%%%%%%%
\section{Equivalence of MBAR and BAR for two states\label{appendix:equivalence-to-bar}}

We start with Eq.~\ref{equation:estimator-of-free-energies}.
%Eq. 13 in the main text.   
For ease of use, we define $\Delta \hat{f} = \hat{f}_2-\hat{f}_1$ and $\Delta u(\x) = u_2(\x) - u_1(\x)$ and $M = \ln N_2/N_1$ Without loss of generalization, since the equations are symmetric, we examine the self-consistent equation for $\hat{f}_1$.
\begin{eqnarray}
\hat{f}_1 &=& -\ln \sum_{j=1}^2 \sum_{n=1}^{N_j} \frac{\exp[-u_1(\x_{jn})]}{\sum\limits_{k=1}^2 N_{k} \, \exp[\hat{f}_{k} - u_{k}(\x_{jn})]} \nonumber \\
%\exp[-\hat{f}_1] &=& \sum_{n=1}^{N_1} \frac{\exp[-u_1(\x_{1n})]}{\sum\limits_{k=1}^2 N_{k} \, \exp[\hat{f}_{k} - u_{k}(\x_{1n})]} + \sum_{n=1}^{N_2} \frac{\exp[-u_1(\x_{2n})]}{\sum\limits_{k=1}^2 N_{k} \, \exp[\hat{f}_{k} - u_{k}(\x_{2n})]} \nonumber \\
1   &=& \sum_{n=1}^{N_1} \frac{\exp[\hat{f}_1-u_1(\x_{1n})]}{N_1\, \exp[\hat{f}_{1} - u_{1}(\x_{1n})] + N_2 \,\exp[\hat{f}_{2} - u_{2}(\x_{1n})]} + \nonumber \\
& &\sum_{n=1}^{N_2} \frac{\exp[\hat{f}_1-u_1(\x_{2n})]}{N_{1} \, \exp[\hat{f}_{1} - u_{1}(\x_{2n})] + N_2 \, \exp[\hat{f}_{2} - u_{2}(\x_{2n})]} \nonumber \\
1   &=& \sum_{n=1}^{N_1} \frac{1}{N_1 + N_2 \,\exp[\Delta \hat{f} - \Delta u(\x_{1n})]} + 
\sum_{n=1}^{N_2} \frac{1}{N_{1} + N_2 \, \exp[\Delta \hat{f} - \Delta u(\x_{2n})]} \nonumber \\
 N_1  &=& \sum_{n=1}^{N_1} \frac{1}{1 + \frac{N_2}{N_1} \,\exp[\Delta \hat{f} - \Delta u(\x_{1n})]} + 
\sum_{n=1}^{N_2} \frac{1}{1 + \frac{N_2}{N_1} \, \exp[\Delta \hat{f} - \Delta u(\x_{2n})]} \nonumber \\
 N_1  &=& \sum_{n=1}^{N_1} \frac{1}{1 + \exp[M + \Delta \hat{f} - \Delta u(\x_{1n})]} + 
\sum_{n=1}^{N_2} \frac{1}{1 + \exp[M + \Delta \hat{f} - \Delta u(\x_{2n})]} \label{equation:intermediate1}
\end{eqnarray}
We make the additional observation that 
\begin{eqnarray*}
 \frac{1}{1 + \exp(x) } - 1 &=& -\frac{1}{1 + \exp(-x)}     
\end{eqnarray*}
which allows us to write~Eq.~\ref{equation:intermediate1} as:
\begin{eqnarray*}
 0 &=& \sum_{n=1}^{N_1} \left[\frac{1}{1 + \exp[M + \Delta \hat{f} - \Delta u(\x_{1n})]}  - 1\right]+ 
\sum_{n=1}^{N_2} \frac{1}{1 + \exp[M + \Delta \hat{f} - \Delta u(\x_{2n})]} \\
   &=& \sum_{n=1}^{N_1} \frac{1}{1 + \exp[M + \Delta \hat{f} - \Delta u(\x_{1n})]} - 
\sum_{n=1}^{N_2} \frac{1}{1 + \exp[-M - \Delta \hat{f} + \Delta u(\x_{2n})]}
\end{eqnarray*}
Which is exactly the equation for BAR presented in Shirts {\em et al.}~\cite{shirts:2003a}.

We now examine the expression for the variance limited to two states. 
When the two thermodynamic states are not identical, the $\bfm{W}$ will have full rank, and the asymptotic covariance matrix can be written as (see Eq.~\ref{equation:covariance-matrix-full-rank} above):
\begin{eqnarray*}
\bfm{\Theta} &=& [(\bfm{W}^\T \bfm{W})^{-1} -\bfm{N} + \bfm{1}_K \bfm{1}_K^{\T}/N]^{-1}
\end{eqnarray*}
where we have from Eq.~\ref{equation:p_weights}
\begin{eqnarray*}
W_{ni} &=& \frac{\exp (\hat{f}_i - u_i(\x_n))}{\sum\limits_{k=1}^K N_{k} \, \exp (\hat{f}_k - u_{k}(\x_n))} 
\end{eqnarray*}
Defining $F$ as the Fermi function, and $X_n = M + \Delta \hat{f} - \Delta u(\x_n)$, then in the case of two states $W_{n1} = N_1^{-1}F(X_n)$ and  $W_{n2} = N_2^{-1}F(-X_n)$.

The matrix $\bfm{W}^\T \bfm{W}$ can then be written as: 
\begin{eqnarray}                                                                         
\bfm{W}^\T \bfm{W}&=&  \sum\limits_{n=1}^{N} \left(\begin{array}{ccc} 
                                                                                N_1^{-2} F(X_n)^2               & N_1^{-1} N_2^{-1} F(X_n) F(-X_n)  \\
                                                                                N_1^{-1} N_2^{-1} F(X_n) F(-X_n) & N_2^{-2} F(-X_n)^2 \\
                                                                         \end{array}\right)
\end{eqnarray}  
If we represent the matrix $(\bfm{W}^\T \bfm{W})_{ij} = a_{ij}$, the determinant $|\bfm{W}^\T \bfm{W}|$ will be $D=a_{11} a_{22} - a_{21} a_{12}$.
The variance of ratios is actually independent of multiplicative factor used in front of $\bfm{1}_K \bfm{1}_K$, as we will show below, so we will use $b$ in place of $1/N$ for generality. The inverse of the covariance matrix is then:
\begin{eqnarray*}                                                                         
\bTheta^{-1} = (\bfm{W}^\T \bfm{W})^{-1} -\bfm{N} + b\, \bfm{1}_N \bfm{1}_N^{\T}  &=&  
                             \left(\begin{array}{ccc} 
                                \frac{a_{22}}{D} - N_1 + b &   -\frac{a_{21}}{D} + b \\
                                -\frac{a_{12}}{D} + b &   \frac{a_{11}}{D} - N_2 + b \\                                                  
                             \end{array}\right)
\end{eqnarray*}
The determinant will then be:
\begin{eqnarray}                                                                         
|\bTheta^{-1}| & = & \frac{1}{D} - \frac{N_2 a_{22} - N_1 a_{11}}{D} + N_1 N_2- b(N_1 + N_2) + b\frac{a_{11}+a_{22} + a_{12} + a_{21}}{D} 
\label{equation:determinant_thetainv}\end{eqnarray}
However, we note that $(\bfm{W}^\T \bfm{W})^{-1}-\bfm{N}$ is singular, and thus the sum of the first three terms in Eq.~\ref{equation:determinant_thetainv} equals zero.  Additionally, because it has kernel $\bfv{1}_K$, it must also satisfy  $a_{22} - a_{21} - N_1 D = 0$ and $a_{11} - a_{12} - N_2 D = 0$.  Because we know by symmetry that $a_{12} = a_{21}$, which we denote by simply $a$, this determinant then becomes:  
\begin{equation*}                                                                         
|\bTheta^{-1}| = \frac{b}{D}\left[a_{11}+a_{22} + a_{12} + a_{21} - (N_1 + N_2) D\right] = \frac{4ab}{D} 
\end{equation*}
We then obtain:
\begin{eqnarray*}                                                                         
\bTheta = [(\bfm{W}^\T \bfm{W})^{-1} -\bfm{N} + b\,\bfm{1}_N \bfm{1}_N^{\T}]^{-1}  &=&  
                             \frac{D}{4ab}\left(\begin{array}{ccc} 
                                \frac{a_{11}}{D} - N_2 + b &   \frac{a}{D} - b \\
                                \frac{a}{D} - b &   \frac{a_{22}}{D} - N_1 + b \\                                                  
                             \end{array}\right)
\end{eqnarray*}
The variance in $f_1 - f_2$ will be $\Theta_{11} + \Theta_{22} - 2\Theta_{12}$, which reduces to:
\begin{eqnarray*}                                                                         
\mathrm{Var}(f_1- f_2)  &=& \frac{D}{4ab}\left[\frac{a_{11} - N_1 D + b D + a_{22} - N_2 D + b D- 2a + 2bD}{D}\right]  \\
                       &=& \frac{1}{4ba}\left[(a_{11} - a - N_2 D) + (a_{22} - a - N_1 D) + 4bD\right] \\
                       &=& \frac{D}{a} 
\end{eqnarray*}
Which is indeed independent of $b \ne 0$.
Since $a_{22} = N_1 + a$ and $a_{11} = N_2 +a$, given $D = a_{11} a_{22} - a^2$ (as noted above), we can find that $D = N_1^{-1} N_2^{-1} (1-Na)$.  We then obtain:
\begin{eqnarray*}
\mathrm{Var}(\hat{f}_1- \hat{f}_2)  &=&\frac{a_{11}a_{22} - a^2}{a}   \\
                                      &=&\frac{1-Na}{N_1 N_2 a} \\
                                      &=&\frac{1}{N_1 N_2 a} - \frac{N}{N_1N_2} \\
                                      &=&\frac{1}{\sum\limits_{i=1}^{N} F(X_n)F(-X_n)} - \frac{N}{N_1N_2} \\                                      
                                      &=&\left[\sum\limits_{i=1}^{N} \frac{1}{2 + 2\cosh(X_n)}\right]^{-1} - \frac{N}{N_1N_2} \\                                      
                                      &=&\frac{1}{N}\left[\left\langle\frac{1}{2 + 2\cosh(M+\Delta \hat{f} -\Delta u(\x))}\right\rangle^{-1} - \left(\frac{N}{N_2} + \frac{N}{N_1}\right)\right]                                      
\end{eqnarray*}
This is the equation for the asymptotic covariance of of free energies given for the BAR method in Shirts \emph{et al.}~\cite{shirts:2003a}
\end{widetext}

%%%%%%%%%%%%%%%%%%%%%%%%%%%%%%%%%%%%%%%%%%%%%%%%%%%%%%%%%%%%%%%%%%%%%%%%%%%%%%%%%%%%%%%%%%%%%%%%%%%%%%
% BIBLIOGRAPHY
%%%%%%%%%%%%%%%%%%%%%%%%%%%%%%%%%%%%%%%%%%%%%%%%%%%%%%%%%%%%%%%%%%%%%%%%%%%%%%%%%%%%%%%%%%%%%%%%%%%%%%
\bibliographystyle{prsty}

\end{document}